\documentclass[eqsecnum,showpacs,amsmath,aps,nofootinbib]{revtex4-1} 
\usepackage{amssymb}
\usepackage{gensymb}
\usepackage{amsmath}

\makeatletter
\renewcommand*{\@fnsymbol}[1]{\ensuremath{\ifcase#1\or *\or \dagger\or
    \ddagger\or 
   \mathsection\or **\or \dagger\dagger
   \or \ddagger\ddagger \else\@ctrerr\fi}}
\makeatother

\begin{document}

\title{Boosted Horizon of a Boosted Space-Time Geometry}
\author{Emmanuele Battista}

\address{Dipartimento di Fisica, Complesso Universitario di Monte S. Angelo, \\
Istituto Nazionale di Fisica Nucleare, Sez. Napoli, Complesso Universitario di Monte S. Angelo,\\
Via Cintia Edificio 6, 80126 Napoli, Italy\\
 E-mail: ebattista@na.infn.it}

\author{Giampiero Esposito} 

\address{Istituto Nazionale di Fisica Nucleare, Sez. Napoli, Complesso Universitario di Monte S. Angelo,\\
 Via Cintia Edificio 6, 80126 Napoli, Italy \\
 E-mail: gesposit@na.infn.it}

\author{Paolo Scudellaro$^*$ and Francesco Tramontano$^{\dagger}$}

\address{Dipartimento di Fisica, Complesso Universitario di Monte S. Angelo, \\
Istituto Nazionale di Fisica Nucleare, Sez. Napoli, Complesso Universitario di Monte S. Angelo,\\
Via Cintia Edificio 6, 80126 Napoli, Italy\\
$^*$ E-mail: scud@na.infn.it \\
$^{\dagger}$ E-mail: tramonta@na.infn.it}

\begin{abstract}
We apply the ultrarelativistic boosting procedure to map the metric of Schwarzschild-de Sitter spacetime into a metric describing de Sitter spacetime plus a shock-wave singularity located on a null hypersurface, by exploiting the picture of the embedding of an hyperboloid in a five-dimensional Minkowski spacetime. After reverting to the usual four-dimensional formalism, we also solve the geodesic equation and evaluate the Riemann curvature tensor of the boosted Schwarzschild-de Sitter metric by means of numerical calculations, which make it possible to reach the ultrarelativistic regime gradually by letting the boost velocity approach the speed of light. Eventually, the analysis of the Kretschmann invariant (and of the geodesic equation) shows the global structure of spacetime, as we demonstrate the presence of a ``scalar curvature singularity'' within a 3-sphere and find that it is also possible to define what we have called ``boosted horizon'', a sort of elastic wall where all particles are surprisingly pushed away. This seems to suggest that such ``boosted geometries'' are ruled by a sort of ``antigravity effect'' since all geodesics seem to refuse entering the ``boosted horizon'' and are ``reflected'' by it, even though their initial conditions are aimed at driving the particles towards the ``boosted horizon'' itself.
\end{abstract}

\maketitle

\section{Introduction}

The sources of gravitational waves can be massless particles moving along a null surface such as a horizon in the case of black holes. In 1971 Aichelburg and Sexl \cite{AS1971} developed a method to describe the gravitational field of a single photon based on the application of a Lorentz transformation to the metric describing a particle at rest, i.e. to the Schwarzschild metric. This method has been called in the literature ``the boost of a metric". Our main attention here will be devoted to the boosting procedure having the Schwarzschild-de Sitter metric as a starting point. Following Refs. \cite{HT1993,ES2007} we can express a de Sitter spacetime in four dimensions as a four-dimensional hyperboloid of radius $a$ embedded in five-dimensional Minkowski spacetime having metric
\begin{equation}
{\rm d}s^{2}_{M}= - {\rm d}Z^{2}_{0}+ {\rm d}Z_{1}^{2}+{\rm d}Z_{2}^{2}+{\rm d}Z_{3}^{2}+{\rm d}Z_{4}^{2},
\end{equation}
with coordinates satisfying the hyperboloid constraint 
\begin{equation}
a^2 = -(Z_{0})^2+ (Z_{1})^{2}+(Z_{2})^{2}+(Z_{3})^{2}+(Z_{4})^{2}.
\label{Z hyperboloid constraint}
\end{equation}
By exploiting the relations between the $Z_i$ ($i=0,1,2,3,4$) coordinates and the spherical static coordinates $(t,r,\theta,\phi)$ (i.e. the coordinates describing the Schwarzschild-de Sitter metric) we can express the Schwarschild-de Sitter metric in the form 
\begin{equation}
{\rm d}s^2 = h_{00}{\rm d}Z_{0}^{2}+ h_{44}{\rm d}Z_{4}^{2}+2h_{04}{\rm d}Z_{0}{\rm d}Z_{4}+ {\rm d}Z_{1}^{2}
+{\rm d}Z_{2}^{2}+{\rm d}Z_{3}^{2},
\label{Z S-dS metric}
\end{equation} 
where $h_{00}, h_{44}$ and $h_{04}$ are functions of the $Z_i$ whose form is not of particular interest for our purposes (see Ref. \cite{ES2007} for details). At this stage, we introduce a boost in the $Z_1$-direction by defining a new set of coordinates independent of $v$, i.e. the $Y_i$ 
coordinates, such that (hereafter $\gamma \equiv  1/ \sqrt{1-v^2} \; $)
\begin{equation}
Z_0 = \gamma \left( Y_0 + v Y_1 \right),    \label{boost 1}
\end{equation}
\begin{equation}
 Z_1 = \gamma \left( v Y_0 + Y_1 \right),
\end{equation}
\begin{equation}
Z_2 = Y_2, \; \;  \; Z_3=Y_3 , \; \;  \; Z_4=Y_4. \label{boost 2}
\end{equation}
Thus, starting from (\ref{Z S-dS metric}) jointly with (\ref{boost 1})--(\ref{boost 2}) we eventually obtain the boosted 
Schwarzschild-de Sitter metric
\begin{eqnarray}
{\rm d}s^2 &=& \gamma^2 \left(h_{00}+v^2\right){\rm d}Y_{0}^{2}+\gamma^2 \left(1+ v^2 h_{00}\right)
{\rm d}Y_{1}^{2}+{\rm d}Y_{2}^{2}+{\rm d}Y_{3}^{2}+h_{44}{\rm d}Y_{4}^{2}  \nonumber \\
&& + 2v\gamma^2 \left(1+h_{00}\right){\rm d}Y_0 {\rm d}Y_1 + 2 \gamma h_{04} {\rm d}Y_0 {\rm d}Y_4 
+ 2 v \gamma h_{04} {\rm d}Y_1 {\rm d}Y_4. \label{boosted metric}
\end{eqnarray}  
On taking the limit $v \rightarrow 1$ of (\ref{boosted metric}), we obtain the boosted Schwarzschild-de Sitter metric in the ultrarelativistic limit
\begin{eqnarray}
{\rm d}s^{2}&=&-{\rm d}Y_{0}^{2}+{\rm d}Y_{1}^{2}+{\rm d}Y_{2}^{2}+{\rm d}Y_{3}^{2}+{\rm d}Y_{4}^{2} \nonumber \\
&+& 4 p \left[-2+{Y_{4}\over a}
\log \left({{a+Y_{4}}\over {a-Y_{4}}}\right)\right]
\delta(Y_{0}+Y_{1})({\rm d}Y_{0}+{\rm d}Y_{1})^{2}, \label{ultrarelativistic boosted metric}
\end{eqnarray}
where the first line describes de Sitter space viewed as a four-dimensional hyperboloid of radius $a$ having equation
\begin{equation}
(Y_{0})^{2}=-a^{2}+(Y_{1})^{2}+(Y_{2})^{2}+(Y_{3})^{2}+(Y_{4})^{2},  \label{hyperboloid constrain}
\end{equation}
embedded into flat five-dimensional space, while the second line of (\ref{ultrarelativistic boosted metric}) describes a shock-wave singularity located on the null hypersurface having equations
\begin{equation}
Y_{0}+Y_{1}=0,
\end{equation}
\begin{equation}
(Y_{2})^{2}+(Y_{3})^{2}+(Y_{4})^{2}-a^{2}=0. 
\end{equation}
The highly singular form of  the distributional metric (\ref{ultrarelativistic boosted metric}) makes the usual spacetime picture no longer valid, but it would
be very interesting to evaluate the effect of these shock-wave singularities on curvature. Therefore, it could be of great physical importance to evaluate the Riemann tensor for this type of geometries, i.e. ``the boosted geometries''. Moreover, as the concept of spacetime curvature is directly related to the geodesic completeness of spacetime, also the analysis of the geodesic equation of the metric (\ref{ultrarelativistic boosted metric}) would be of great relevance for the description of the topological features of spacetime, in particular the symmetries and the singularities.

\section{The ``Boosted'' Riemann Curvature Tensor and The Kretschmann Invariant} 

The high degree of difficulty arising in the relations describing ``the boosted geometry'' we are handling is such that the only way we had to compute the Riemann-Christoffel symbols and the Riemann tensor after the boost was represented by numerical calculations. In this way we can evaluate the behaviour of spacetime curvature also in the ultra-relativistic regime, which is the one we are mainly interested in, by letting the velocity defined by the boost relations (\ref{boost 1})--(\ref{boost 2}) approach gradually the speed of light. Thus, starting from the apparently $5 \times 5$ metric (\ref{boosted metric}), we can restore the usual four-dimensional form of the metric by exploiting the hyperboloid constraint (\ref{hyperboloid constrain}). By virtue of this condition we obtain the manifestly four-dimensional form of the boosted metric (\ref{boosted metric}), which can be expressed by the relations \cite{BEST}
\begin{equation}
g_{11} =  \dfrac{\gamma^2 \left(h_{00}+v^2\right)}{\sigma} Y_{1}^{2}+\gamma^2 \left(1+v^2h_{00}\right)+\dfrac{2v\gamma^2  
\left(1+h_{00}\right)}{\sqrt{\sigma}} Y_1 \label{g11} ,
\end{equation}
\begin{equation}
g_{jj} = \dfrac{\gamma^2 \left(h_{00}+v^2\right)}{\sigma} Y_{j}^{2} +1,  \; \; \; (j=2,3)
\end{equation}
\begin{equation}
g_{44} = \dfrac{\gamma^2 \left(h_{00}+v^2\right)}{\sigma} Y_{4}^{2} +h_{44}+ \dfrac{2 \gamma h_{04}}{\sqrt{\sigma}} Y_4 ,
\end{equation}
\begin{equation}
g_{1j} = \dfrac{\gamma^2 \left(h_{00}+v^2\right)}{\sigma} Y_{1}Y_{j} + \dfrac{v \gamma^2 \left(1+h_{00} \right)}{\sqrt{\sigma}}Y_{j} + \delta_{4j} \left(  \dfrac{\gamma h_{04}}{\sqrt{\sigma}}+v \gamma h_{04}   \right),  \; \; (j=2,3,4),
\end{equation}
\begin{equation}
g_{2j} = \dfrac{\gamma^2 \left(h_{00}+v^2\right)}{\sigma} Y_{2}Y_{j}+ \delta_{4j} \left(   \dfrac{\gamma h_{04}}{\sqrt{\sigma}} Y_2 \right), \; \; \; (j=3,4),
\end{equation}
\begin{equation}
g_{34} = \dfrac{\gamma^2 \left(h_{00}+v^2\right)}{\sigma} Y_{3}Y_{4} + \dfrac{\gamma h_{04}}{\sqrt{\sigma}} Y_3, \label{g34}
\end{equation}
where $\sigma(Y_{\mu }) \equiv -a^{2}+Y_{1}^{2}+Y_{2}^{2}+Y_{3}^{2}+Y_{4}^{2}$. Having obtained the formulas (\ref{g11})--(\ref{g34}), we can evaluate the Riemann-Christoffel symbols and consequently the Riemann 
curvature tensor of (\ref{boosted metric}) by using the familiar relations of classical general relativity. In order to study the features of the Riemann curvature of spacetime described by the metric (\ref{ultrarelativistic boosted metric}), 
we therefore decided to plot the Kretschmann invariant at different values of boost velocity $v$ and study the geodesic equation
\begin{equation}
\ddot{Y}^{\mu}(s) + \Gamma^{\mu}_{\; \nu \lambda} \dot{Y}^{\nu}(s) \dot{Y}^{\lambda}(s)=0, \label{geodesic eq}
\end{equation} 
$s$ being the affine parameter of the geodesic having parametric equation $Y^{\mu}=Y^{\mu}(s)$.
From the analysis of the Kretschmann invariant we found \cite{BEST} that it is not defined unless the inequality 
(hereafter, numerical values of $Y$ coordinates have downstairs indices, to be consistent with the notation adopted so far)
\begin{equation}
(Y_{1})^2+(Y_{2})^2+(Y_{3})^2+(Y_{4})^2 > a^2, \label{singularity kretschmann}
\end{equation}
is satisfied. Hence, we see that the hyperboloid constraint, i.e. condition (\ref{hyperboloid constrain}), allows us to define a 3-sphere of 
radius $a$ and having equation $(Y_{1})^2+(Y_{2})^2+(Y_{3})^2+(Y_{4})^2=a^2$ where the Kretschmann invariant  is not defined. As we have discovered by the joint analysis of the geodesic equation (\ref{geodesic eq}) that the Kretschmann invariant blows up along a geodesic, this 3-sphere represents a ``scalar curvature singularity''. This peculiar feature of our ``boosted spacetime geometry'' is indeed obvious 
if we look at formulas (\ref{g11})--(\ref{g34}), as here the quantities $\sigma$ and $\sqrt{\sigma}$ always appear at the denominator of 
the expressions of the metric tensor $g_{\mu \nu}$, which means that the metric is defined only if the inequality 
(\ref{singularity kretschmann}) holds. Moreover, if we interpret $Y_0$ as the time coordinate (see (\ref{boost 1})), we can view 
(\ref{singularity kretschmann}) as a condition on time.

Another interesting feature of ``boosted geometries'' that we have found \cite{BEST} consists in the presence of a sort of barrier surrounding the  
3-sphere, which we may call ``boosted horizon'', in the sense that all geodesics, despite maintaining their completeness condition,  
are surprisingly pushed away from it. We have also discovered that the extension of the ``boosted horizon'' depends only on the boost velocity $v$ and not on the particle's initial velocity. Since we have found that all geodesics are complete, according to standard definitions of general relativity the ``boosted horizon'' 
is not a singularity but it seems to be a sort of elastic wall which is hit by all particles before they get away. In the ultrarelativistic regime ($v$ = 0.9999) the “antigravity effects” are still present but the position of the boosted horizon tends to that of the singularity 3-sphere \cite{BEST}.
 
\section{Conclusions}

The most interesting aspects of our research consist in the discovery of this elastic wall which we have proposed to call ``boosted horizon'' where all solutions of (\ref{geodesic eq}) ``refuse'' to be attracted, regardless of their initial velocity, and, moreover, in the presence of the singularity 3-sphere, which always lies inside the ``boosted horizon'', representing for the ``boosted geometry'' a ``scalar curvature singularity''. We suppose that ``antigravity effects'' may result from the term $\Lambda = 3/a^2 >0$ occurring in the Schwarzschild-de Sitter metric (a positive cosmological constant $\Lambda$ represents a repulsive interaction), while the ``scalar curvature singularity'' might be related to the presence of a more exotic object, i.e. a firewall \cite{AMPS,Braunstein2013,Braunstein2015}, which can be a possible solution to an apparent inconsistency in black hole complementarity \cite{STU,SHW}.

\end{document}